\documentclass[12pt, english, headsepline,a4paper]{article}
\usepackage{graphicx} 
\usepackage[a4paper, total={6in, 8in}]{geometry}
\usepackage{amsmath}
\usepackage[numbers]{natbib}
\usepackage[utf8x]{inputenc}
\usepackage{mathptmx} 
\usepackage{hyperref}
\usepackage{tabto}
\usepackage{subcaption}
\usepackage{longtable}
\usepackage{array}
\newcolumntype{C}[1]{>{\centering\arraybackslash}p{#1}}
\newcolumntype{L}[1]{>{\raggedright\arraybackslash}p{#1}}
\hypersetup{
    colorlinks,
    citecolor=black,
    filecolor=black,
    linkcolor=black,
    urlcolor=black
}
\title{}
\date{}

\begin{document}

\newgeometry{left=1.5cm,right=1.5cm,top=2.5cm,bottom=2.5cm}
\pagenumbering{gobble}

\begin{center}

	\vspace{2.5 cm}
	{\fontsize{30 pt}{30 pt}\selectfont \bfseries  ETHER\par}
    \hspace{4 cm}
	\vspace{4 cm}
	
	\textit{Efficient Tool for THermodynamics Exploration via Relaxations}
	
	\vspace{2 cm}
	
	\text{developed by} \\ \vspace{2 cm}
	\textbf{Dr. Mukesh Kumar Sharma}\\ \vspace{0.5 cm}
	\textit{PhD, Indian Institute of Technology Roorkee, India}

	\vspace{5 cm}
	
	\begin{figure}[h]
		\centering
	\end{figure}
	\vspace{0.5 cm}
	
\end{center}

\newpage

\section*{Preface}

Monte Carlo (MC) simulations are powerful tools for validating analytical methods and solving large, complex systems where analytical approximations become challenging. Motivated by these compelling features, I have developed a simulation package named ETHER (Efficient Tool for THermodynamics Exploration via Relaxations), designed to uncover temperature dependent phenomena in magnetic materials. Using ETHER, users can perform extensive MC simulations for various spin systems to study phase transitions, critical points, and complex magnetic structures.

The initial development of ETHER began during my Ph.D. tenure (2017 to 2023) at the Indian Institute of Technology Roorkee, India, with the objective of investigating magnetic materials and their role in multiferroic properties. The simulation package is primarily written in FORTRAN, with core scientific calculations implemented in well structured and easily understandable subroutines. In addition, ETHER supports parallel computation in hybrid mode using both Open Multi Processing (OpenMP) and Message Passing Interface (MPI) protocols, significantly improving computational efficiency. Over time, the package has been rigorously tested and several bugs have been addressed.

Accurate MC simulations for complex spin Hamiltonians require precise information about the spin lattice and neighboring magnetic ions. Incorrect lattice definitions can lead to unexpected or erroneous results. While defining crystal lattices via space group symmetry operators is straightforward, handling doped or antisite disordered systems is more tedious. Doping and disorder are typically introduced at the unit cell level, but achieving correct statistical representations requires the construction of larger supercells \cite{Landau1976, Landau1978}. However, using large supercells increases computational demands and time, especially when studying disorder effects.

 ETHER overcomes this limitation. Unlike conventional methods where doping or disorder is confined to the unit cell, ETHER allows the implementation of desired concentrations directly in a supercell. This feature gives users the flexibility to perform MC simulations for varying concentrations within a fixed supercell size, thereby saving computational resources and enabling systematic exploration of disorder effects. One can also generate fully randomized doping configurations by constructing special quasirandom structures (SQS), which provide representative atomic arrangements for a given composition and enable realistic modeling of disordered systems. Several software packages are available for this purpose, such as the Alloy Theoretic Automated Toolkit (ATAT)\cite{van2002alloy} and the sqsgenerator code\cite{lebeda2025simplysqs}, which arrange dopant atoms in a supercell so that the statistical pair correlation functions over several neighbour distances closely match those of a perfectly disordered system.

For a given magnetic material, ETHER constructs the spin lattice network internally using the input structure file `structure.vasp'. This file, similar in format to POSCAR files used in VASP \cite{vasp1, vasp2}, can be easily generated using visualization software such as VESTA \cite{vesta}. Exchange interactions, which define how a magnetic ion interacts with its neighbours, are provided via the `\textit{j\_exchange}' input file. ETHER identifies neighbouring ions either by their names or by the bond lengths specified in this file. These neighbour lists can be confirmed and visualized using tools such as XCrySDen\cite{xcrysden} or VESTA. Additionally, several other input files help the program capture detailed information about the spin system. Upon completion of MC simulations, ETHER generates multiple output files, including final spin states that can be visualized using VESTA. The package also provides post processing tools to help users analyze the simulation data effectively.

The primary objective of developing ETHER is to offer researchers a flexible, user-friendly tool for temperature-dependent Monte Carlo simulations of complex magnetic materials. I hope users will find the code intuitive and easy to work with. Detailed instructions on installation, input/output formats, and visualization methods are provided in this user guide. Researchers are encouraged to contribute to the ongoing development of ETHER.


\newpage

\newpage
\tableofcontents
\newpage
\pagenumbering{arabic}

\section{Code Installation}
The $ETHER$ source code can be downloaded from \textit{https://github.com/mukkelian/Ether} \cite{ether_source}. \textit{Users} are required to have gcc or Intel based compilers on their system to install the code \textit{ETHER}. After downloading ETHER source code, please follow the given steps for installation procedure in the terminal,\\

\fbox{\textit{COMMAND} $\Longrightarrow$ cd $<$path/to/ether$>$}\vspace{0.5 cm}

\fbox{\textit{COMMAND} $\Longrightarrow$ make all}
\vspace{0.5 cm}

Once the installation is done, an executable `\textit{ether}' will be generated in the main directory. The choice of compilers can be configured in the \textit{make.sys} file located in the same directory. Users can copy the executable to the directory where MC simulations will be performed. A script file, \textit{run\_ether.sh}, is also provided in \textit{$<$path/to/ether$>$/inputfiles/} for submitting jobs on multi node HPC clusters to enable scalable performance. The \textit{ether} executable must be present in the directory from which \textit{run\_ether.sh} is executed.\\
User can follow the step for submitting their jobs on multi-node HPCs.\\

\fbox{\textit{COMMAND} $\Longrightarrow$ bash \textit{run\_ether.sh}}

\section{Hamiltonian}

We have implemented the following form of Hamiltonian for evaluating the thermodynamic properties of any spin system.

\begin{equation}\label{xyz}
    H = \sum_{<i,j>}^{}\left(J_{xx}^{ij} S_{x}^{i} S_{x}^{j} + J_{yy}^{ij} S_{y}^{i} S_{y}^{j} + J_{zz}^{ij} S_{z}^{i} S_{z}^{j}\right) + \sum_{i}^{}[D^{i}_{x} (S^{i}_{x})^2 + D^{i}_{y} (S^{i}_{y})^2 + D^{i}_{z} (S^{i}_{z})^2]  - g\mu_B\sum_{i}^{}[H_xS^{i}_{x}+H_yS^{i}_{y}+H_zS^{i}_{z}]
\end{equation}

Classical spin vector: \tab$\vec S$ [$S_x$, $S_y$, $S_z$]

Exchange interaction: \tab$J$ [$J_{xx}^{}$, $J_{yy}^{}$, $J_{zz}^{}$]

Magnetic field: \tab$\vec H$ [$H_x$, $H_y$, $H_z$]

Single ion anisotropy vector: \tab$\vec D$ [$D_x$, $D_y$, $D_z$]

g-factor: \tab$g$

Bohr magneton: \tab$\mu_B$\\

The \textit{First}, \textit{Second} and \textit{Third} term represents the classical XYZ spin model, single ion anisotropy, and Zeeman terms, respectively. $< >$ stands for nearest neighbours and here $i$ represents the $i$-th spin site. Here if $J>0$ then interaction will be anti-ferromagnetic else if $J<0$ then system will have ferromagnetic interactions whereas spins prefers to align towards the direction where negative component of $\vec D$ is present. 
This Hamiltonian (eqn.\ref{xyz}) can also be converted into classical Heisenberg model at the condition when $J_{xx} = J_{yy} = J_{zz} = J$ or into any anisotropic model like XXZ  ($J_{xx} = J_{yy}$) or, XY model ($J_{xx} = J_{yy} = J,  J_{zz} = 0$). The choice of model either Ising or XYZ can be provided by the user in the input file `$in.ether$' (see \textit{MODEL} tag in Inputs section 6.1).

If logical variable $PARA$ (\textit{see: Units}) is set to $.true.$, for a given $|J_{para}| > 0$, then $Ether$ considers the following effective Hamiltonian,\\

\begin{equation}\label{para}
	\tilde{H} = \dfrac{H}{|J_{para}|}
\end{equation}

\section{Units}
$ETHER$ by default takes:\\

\textbf{IF} [PARA = .FALSE., $see$ $in.ether$]\\

\begin{tabular}{l}
\hspace{1 cm} $J$ in terms of milli-electron-Volts (meV)\\

\hspace{1 cm} $\vec D$ in terms of meV\\

\hspace{1 cm} $\vec H$ in Tesla\\

\hspace{1 cm} $\mu_B$ in eV/Tesla (internally fixed)\\

\hspace{1 cm}  Bond lengths in terms of Angstrom (\AA)\\

\hspace{1 cm}  Temperature(T) range in terms of Kelvin (K)\\  
\end{tabular}\\

\textbf{ELSE} [PARA = .TRUE. with given J$_{para}$, $see$ $in.ether$]\\

\begin{tabular}{l}
\hspace{1 cm} $J$ as unit less\\

\hspace{1 cm} $\vec D$ as unit less\\

\hspace{1 cm} $\vec H$ as unit less\\

\hspace{1 cm} $\mu_B$ = 1 \\

\hspace{1 cm}  Bond lengths in terms of Angstrom (\AA)\\

\hspace{1 cm}  Temperature range in terms of $k_BT/J$ such that $J$ = $|J_{para}|$\\
\end{tabular}\\


\section{Calculating Observables (A)}

The framework of the conventional Monte Carlo technique is built on the basis of statistical mechanics. Where, we calculate any thermodynamic property by taking its average over all possible states. Suppose, $A$ is our thermodynamical observable in which we are interested to calculate, and $A_s$ denotes the observable when the system is in state $s$ with energy $E_s$, where $s$ is the state of system in the sample space. Then the expectation ($<$A$>$) of observable $A$ can be defined as,
\begin{equation}\label{average}
	\left <A \right > = \dfrac{1}{Z}\sum_{s}^{}A_se^{-\beta E_s}
\end{equation}
where $Z$ is the partition function and is defined as $\sum_{s}^{}A_se^{-\beta E_s}$ such that it covers all states through which the system undergoes, $\beta = \frac{1}{k_BT}$. Once the exact value of $Z$ is known, $<$A$>$ can be derived exactly. However, it is difficult to know the exact value of $Z$\cite{Luijten}. In 1953, Nicholas Metropolis and his co-workers\cite{metropolis1953} deduced an astonishing method to evaluate the expectation value of targeted thermodynamic observable without evaluating $Z$ and thus empowered the MC simulations that we currently know now.

In the MC simulation, the basic idea is to obtain a series of states, called \textit{Markov chain}, in which each state, say $s_i$, only depends on its immediately preceding state $s_{i-1}$. In this process, the inclusion of new states can only be possible if it satisfies the specific transition probability criteria. Suppose a system at state $s_i$ with energy $E_i$ such that Boltzmann factor $p_i \propto e^{-\beta E_i}$, and our trial new configuration $s_j$ with energy $E_j$ having Boltzmann factor $p_j$. Then the acceptance or rejection of this new trial configuration will be done according to the following condition
\begin{equation}\label{MCcondition1}
P_{ij} = 
\begin{cases}
	\text{$e^{-\beta(E_{j}-E_{i})}$}	& \text{if  $E_j\,>\,E_i$}\\
	1,											& \text{if $E_j\, \leq \, E_i$}\\
\end{cases}
\end{equation}
or, often above condition can also be written as,
\begin{equation}\label{MCcondition2}
	P_{ij} = min[e^{-\beta \Delta_{ij}}, 1]\\
\end{equation}

Here, $P_{ij}$ is the acceptance probability, and  $ \Delta_{ij} = E_{j}-E_{i} $ is the energy difference. If trail state $s_j$ is accepted, the associated configuration will be saved and considered as the new member of the series, and if it is rejected, the next member will be again $s_i$. Generally, the criteria given in equation \ref{MCcondition1} is implemented using random number $r\in [0, 1) $ and the new state will be accepted only if $r< P_{ij}$.

Following the above condition, suppose we have a list of $n$ states $s_1$, $s_2$, ..., $s_n$ and the calculated observable $A_i$ is associated with the state $s_i$. Then the thermodynamic property $A$, can be estimated by simply averaging the $A_i$, i.e.,
 
\begin{equation}\label{observable}
	\left<A\right> \approx \dfrac{1}{n} \sum_{i=1}^{n}A_i
\end{equation}

ETHER can calculate the following observables at temperature (T),\\

\textbf{Total internal energy}$^*$ per site,
\begin{equation}\label{eng}
    E(T) = \dfrac{1}{2N}\langle H \rangle
\end{equation}

\textit{$^*$Integer 2 in denominator is added for removing the double counting}\\

 \textbf{Specific heat} (\textit{dimensionless}) per site,
\begin{equation}\label{sh}
    \dfrac{C(T)} {k_B} = \beta^2\left(\left< E^2 \right> - \left< E \right>^2\right)N  \text{;}  \hspace{2 cm}\beta = \dfrac{1} {k_BT}
\end{equation}

 \textbf{Magnetization}$^*$ per site,
 \begin{equation}\label{mag}
 M = \dfrac{1}{N}\langle |\sum^{N}_{i=1}c_i\vec S_i|\rangle
 \end{equation}

 \textit{$^*$ $c_i$ is the staggered coefficient (see section 6.4 for more details)}\\
 
\textbf{Susceptibility} per site,
\begin{equation}\label{chi}
    \chi (T) = \beta\left(\left< M^2 \right> - \left< M \right>^2\right)N
\end{equation}

\textbf{Structure factor }\cite{kinzel1982, saito1981},
\begin{equation}\label{sf}
S_f(q) = \frac{1}{N}\sum_{r,\,r'} \left <\,S_{r} S_{r'} \right > e^{iq(r-r')} 
\end{equation}

\textbf{Cumulant} \cite{binder1981_1, binder1981, landau1983},
\begin{equation}\label{Cuml}
U_A = 1 - \dfrac{1}{3} \dfrac{\left< A^4 \right>}{\left< A^2 \right>^2} 
\end{equation}

\section{Standard Deviation and Error Estimation of Observable}
For an observable $A$, if $\langle A \rangle$ is calculated $m$ times repeatedly such that obtained values are $\langle A \rangle_1$, $\langle A \rangle_2$,...,$\langle A \rangle_m$ and their evaluated mean is $\langle A \rangle_{mean}$,

\begin{equation}\label{avgA}
\langle A \rangle_{mean} = \frac{1}{m} \sum_{i=1}^m \langle A \rangle_i
\end{equation}
then,\\

\vspace{1pt}

 \textbf{Standard deviation} in calculated $A$ is:

\begin{equation}\label{sv}
\sigma_A = \sqrt{\frac{1}{m - 1} \sum_{i=1}^m (\langle A \rangle_i - \langle A \rangle_{mean})^2}
\end{equation}

\textbf{Standard error (uncertainty)} is:
\begin{equation}\label{se}
\Delta A = \frac{\sigma_A}{\sqrt{m}}
\end{equation}

\noindent thus $A$ can be reported as:
$A = \langle A \rangle_{mean} \pm \Delta A$

\section{Inputs}
\subsection{$in.ether$}
`$in.ether$' file contains the necessary input information based on which $ETHER$ will build-up the internal framework and initiate the MC simulation accordingly. User can provide the following \textit{Tags} for initiating the MC simulation:\\

\begin{longtable}{|C{2.5cm}|L{11cm}|C{3cm}|}
\hline
\rule{0pt}{0.9cm}\textbf{Tags} & \textbf{Description} & \textbf{Default} \\
\hline
MODEL & Defines the choice of model considered during MC simulation. \newline (\textit{Ising/XYZ}) & Ising \\
\hline
MCS & Monte Carlo steps per spins (MCS)\newline (Total MCS for simulation, Total MCS for equilibration process) & 5000 3000 \\
\hline
TEMP & Sets the temperature range. \newline\textit{Note: Simulation will start from higher temperature.} \newline (T$_{Start}$, T$_{End}$, T$_{interval}$) & 100 10 5\\
\hline
SPIN & Spin values of each ion present in the $structure.vasp$ file. \newline \textit{Note: For non magnetic ion, kindly assign 0.0} &\\
\hline
SPECIES & Symbol of species for which MC simulations is intended to perform. & \\
\hline
SC & Size of supercell along x, y, and z direction. \newline (\textit{X, Y, Z}) & 2 2 2\\
\hline
STG & Logic for staggered magnetization. If .TRUE. then $ETHER$ will require the mandatory `$staggered$' file for further inputs, \textit{see: staggered}. \newline (\textit{Logic}) & .FALSE.\newline $c_i$ = 1 [\textit{eqn.} \ref{mag}]\\
\hline
BC & Boundary conditions (c: Closed, o: Open) along x, y, z direction \newline (X, Y, Z) & c c c\\
\hline
REPEAT & For any observable, repeat the same MC simulations $m$ times for better results, \textit{check equation} \ref{avgA}. & 10 \\
\hline
ANGLE & To update the direction of spin vector.\newline 
$Note$: If this logic is .TRUE. then spins will be updated according to $S_x = \sqrt{1-S_z^2}\cos\phi$, $S_y = \sqrt{1-S_z^2}\sin\phi$, and $-1\le S_z \le+1$ such that $\phi = \phi_{old} + \phi^{'}\delta \,(0\le \delta \le 1)$ where $\phi_{old}$ is the older {$\phi$}. Otherwise, new spins follows the George Marsaglia\cite{Marsaglia} rule. \newline (\textit{logic, $\phi$}) & .FALSE. 5\\
\hline
COA & Calculate observables at each defined MCS. & 10 \\
\hline
ZEEMAN & Zeeman logic with external magnetic field. \textit{see equation} \ref{xyz}. \newline (\textit{logic, $H_x$, $H_y$, $H_z$}) & .FALSE. 0 0 0\\
\hline
G\_FACTOR & g\_factor value & 2 \\
\hline
SIA & Single ion anisotropy logic. \textit{see equation} \ref{xyz}. \newline (\textit{logic}) & .FALSE.\\
\hline
PARA & Parameter logic with J$_{para}$ in meV. \textit{see equation} \ref{para}. \newline (\textit{logic, J$_{para}$}) & .FALSE. 1.0\\
\hline
OVRR & Logic for performing over-relaxation (OVRR) \cite{ovrr1, ovrr2} method applied on total (over\_para $\times$ total\_magnetic\_lattices) lattices after each ovrr\_MCS steps. \newline (\textit{logic, over\_para, ovrr\_MCS}) & .FALSE. 0 0\\
\hline
SEED & Seed number for random number generation & 1992\\
\hline
NBDFC & Neighbourhood finding criteria. \newline (\textit{nbd\_finding\_criteria}) & 10$^{-5}$ \\
\hline
\end{longtable}

\subsection{$j\_exchange$}
The `$j\_exchange$' file contains the list of magnetic exchange interactions (J) acting between magnetic ions. Since $ETHER$ recognizes the $J^{AB}$ interactions either from the species symbols (A, B) of the ions or from the corresponding bond lengths ($d_{AB}$), $users$ must carefully provide the correct $J$, $d$, and $species \,\,symbol$ information when preparing this file.

\vspace{10 pt}

\noindent Let's assume, we have two magnetic ions (A, B) present in the unit cell and $d_{AB}$, $d_{AA}$, $d_{BB} $ are the connected bond lengths$^*$ between magnetic ion pair A-B, A-A, and B-B respectively such that $d_{AA}$ = $d_{BB} = 7 \AA$ and $d_{BB} = 5 \AA$ then, following is the expected file format for making $j\_exchange$ file,
\vspace{14 pt}

\noindent ================================================\\
\begin{tabular}{l l}
2 & ! Number of distinct bonds (2: $d_{AB}$  and $d_{AA}$ = $d_{BB}$)\\
5.0 1 & ! Bond length $d_{AB}$, number of similar bonds (1)\\
A B -1.0 2.0 3.0 & ! Ion A, Ion B, $J_{xx}^{AB}$, $J_{yy}^{AB}$, $J_{zz}^{AB}$\\

\noindent7.0 2 & ! Bond length $d_{AA}$, number of similar bonds (2)\\
A A 1 1 1 & ! Ion A, Ion A, $J_{xx}^{AA}$, $J_{yy}^{AA}$, $J_{zz}^{AA}$\\
B B 2 3 4.0 & ! Ion B, Ion B, $J_{xx}^{BB}$, $J_{yy}^{BB}$, $J_{zz}^{BB}$\\
\end{tabular}

\noindent ================================================\\
\textit{Note: J values presented above are just for an example. \newline $^*$ Use VESTA\cite{vesta} software to see how many distinct bonds are present among the magnetic ions.}\\

\subsection{$single\_ion\_anisotropy$}
For calculations including single ion anisotropy (SIA), the flag `SIA = .TRUE.` must be set. Users must then provide the components of the SIA vector ($D_x$, $D_y$, $D_z$) for the desired species present in the $structure.vasp$ file into the $single\_ion\_anisotropy$ input file.

\vspace{10 pt}
\noindent The format of the $single\_ion\_anisotropy$ file is given below :

\noindent ================================================\\
\begin{tabular}{l l l}
$Species_i$ & $D_x^i$ $D_y^i$ $D_z^i$ & ! For Species(i) in unit cell\\
$Species_{i+1}$ & $D_x^{i+1}$ $D_y^{i+1}$ $D_z^{i+1}$ & ! For Species(i+1) in unit cell\\
.           .& &\\
.           .& &\\
.           .& &\\
\end{tabular}\\
\noindent ================================================

\subsection{$staggered$}
If the system under study is antiferromagnetic, evaluating the magnetization or its derivatives, such as susceptibility or the cumulant, becomes challenging because these quantities depend on the average sum of spins, which becomes zero or negligibly small below the critical temperature.

\textit{Staggered magnetization}, defined as the difference between the magnetizations of two antiferromagnetically coupled magnetic sublattices, can recover the information that is absent in the total magnetization. For evaluating the staggered magnetization, $users$ must set the variable STG = .TRUE. in the $in\_ether$ input file and provide the staggered coefficients ($c_i$, \textit{see equation \ref{mag}}) with values of +1, -1, or other real numbers in the `$staggered$' file. $ETHER$ will then construct two distinct magnetic sublattices based on the sign of $c_i$ and perform the MC simulations accordingly. 
\newline
\textit{Note:} Performing staggered calculations does not affect the observables calculated from the total internal energy $E$.

\vspace{10 pt}
\noindent The format of the $staggered$ file is given below:
\vspace{10 pt}
\newline
\newline
\newline
\newline
\noindent 
============================================\\
\begin{tabular}{l l}
$c_i$ & ! For atom(i) in unit cell\\
$c_{i+1}$ & ! F or atom(i+1) in unit cell\\
.           . &\\
.           . &\\
.           . &\\

\end{tabular}\\
\noindent ============================================\\

\section{Outputs}
The following sub-sections gives the information and the structure of the generated output files after running $ETHER$ code.

\subsection{$<$tt.tttt$>$spK.xsf}
\begin{figure}[h!]
\centering
\includegraphics[width=0.5\textwidth]{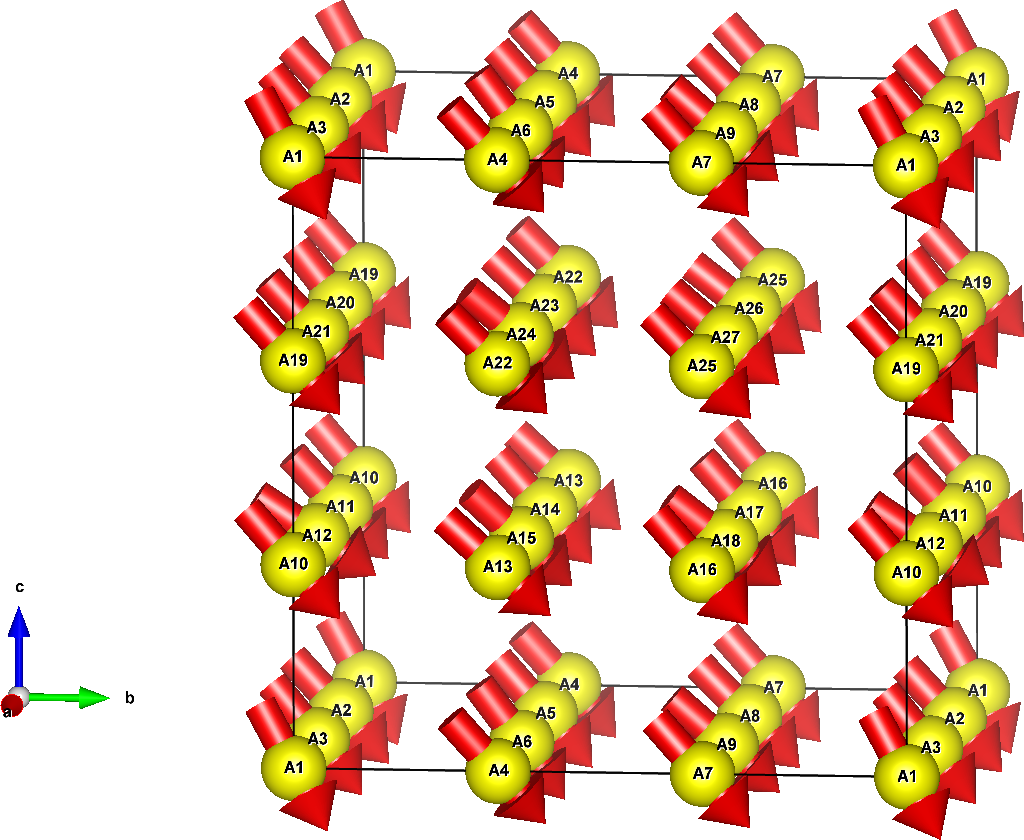}
\caption{The ground state arrangement of a simple cubic system of size 3$\times$3$\times$3 after running Monte Carlo simulations using $Ether$ code.\newline \textit{After running the simulation, results can be obtained at location:}\\ /examples/SimpleCubic/333/spins/2.0000spK.xsf}
\label{spin_config}
\end{figure}

Files named `$<$tt.tttt$>$spK.xsf' are output files generated during the MC simulations at temperature $<$tt.tttt$>$ and contain the calculated final ground spin states. After the simulation is completed, all files ending with $*spK.xsf$ can be found in the $/spins$ directory. Users can therefore analyze the spin configuration at any temperature using the VESTA program \cite{vesta}. In this section, results are shown for the special case where $J$ is considered only for first nearest neighbours. Figure \ref{spin_config} illustrates the ground spin configurations calculated at a temperature of 2 K for a supercell size of $3\times3\times3$.


\subsection{$nbd.dat$}
\begin{figure}[h!]
\centering
\fbox{\includegraphics[width=1.0\textwidth]{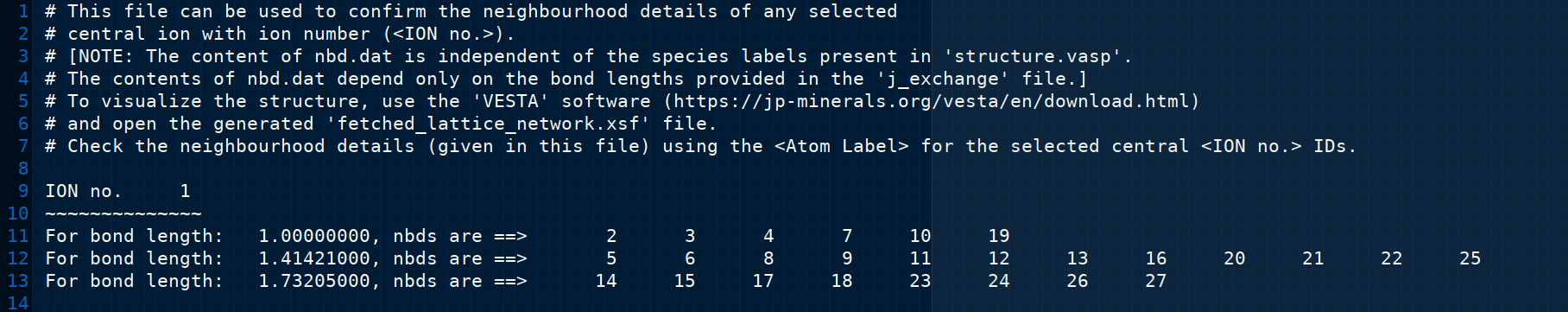}}
\caption{ Showing the first few lines of nbd.dat file.}
\label{nbd_ss}
\end{figure}

The list of nearest neighbourhood with respect to any central ion in a supercell can be found in this file. The format of $nbd.dat$ can be understood as follows (see Figure \ref{nbd_ss}): \newline

The commented lines (starting with the character \#) explain the procedure for visualizing the neighbouring ion details. If the `structure.vasp' file contains different magnetic ion species, users are advised to rename them with a single species label in order to visualize the neighbouring ions of the central atom along with their corresponding IDs (ION no. $<$ID$>$). It should be noted that the information in \textit{nbd.dat} is independent of the species labels and is generated solely based on the bond length information provided in the \textit{j\_exchange} file.


\begin{figure}[h]
\centering
\fbox{\includegraphics[width=1.0\textwidth]{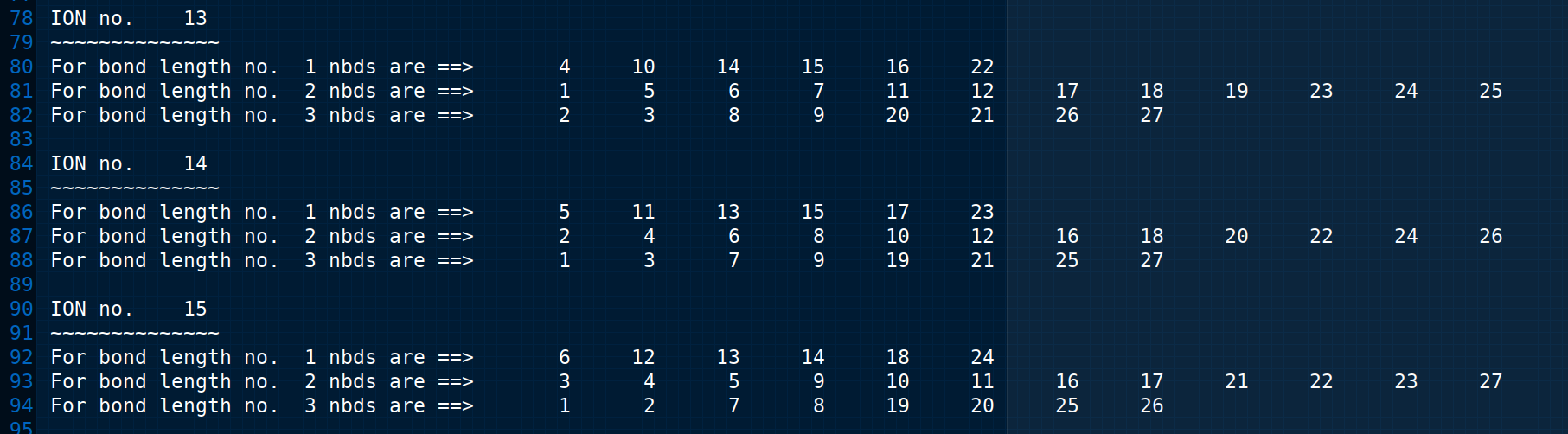}}
\caption{Shows the contents of $nbd.dat$ file, which comprises of nearest neighbour details for the \textbf{\textit{13-th, 14-th, and 15-th} ions} marked in Figure \ref{spin_config}. This file can be used to verify whether the nearest neighbour atoms were correctly identified.\newline\textit{After running the simulation, results can be obtained at given location:}\\ /examples/SimpleCubic/333/data/nbd.dat}
\label{nbd.dat}
\end{figure}

\begin{figure}[h!]
\centering
\includegraphics[width=0.6\textwidth]{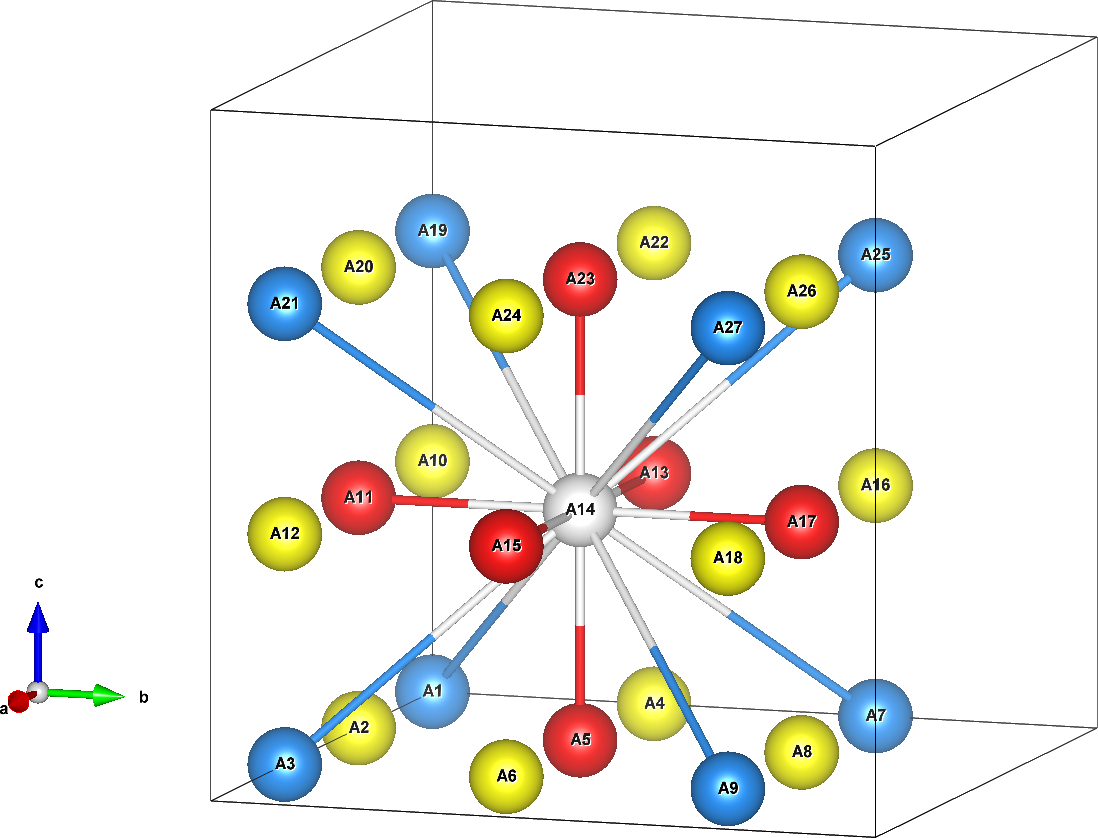}
\caption{Visualization of the generated nbd details of \textbf{ION no.} 14. from the file, `\textit{fetched\_lattice\_network.xsf}' using VESTA. Here 1$^{st}$ , 2$^{nd}$  and 3$^{rd}$ nearest neighbour are represented by red , blue, and yellow spheres, respectively.}
\label{14th_ion}
\end{figure}

In figure \ref{nbd.dat}, we have shown one such example for the \textbf{ION no.} 14 present in a simple cubic lattice (See Figure \ref{spin_config} for the lattice) with supercell size 3$\times$3$\times$3. The ions till the third nearest neighbours are considered (See Figure \ref{14th_ion}). $Users$ are advised to rerun the calculation in the given directory to reproduce the nearest neighbourhood (nbd) in $nbd.dat$ and further analyse it. \\

In this file, the number of distinct bond lengths specified in $j\_exchange$ file defines the number of nbds. Depending on the requirement, $user$ can extend to the higher nearest neighbours by specifying the corresponding distinct number of bonds, their lengths  and their associated connecting ions in the $j\_exchange$.


\subsection{$gss.dat$}
The ground spin states file ($gss.dat$) contains information about the final spin configurations obtained at each temperature during the simulation. It also includes essential details related to the simulation box used during the initialization of the Monte Carlo simulation. Users are strongly advised not to edit this file, as it contains core data required by the post processing tools available in ETHER's utility folder.

\subsection{$moment\_vector.dat$}
\begin{figure}[h!]
\centering
\fbox{\includegraphics[width=1.0\textwidth]{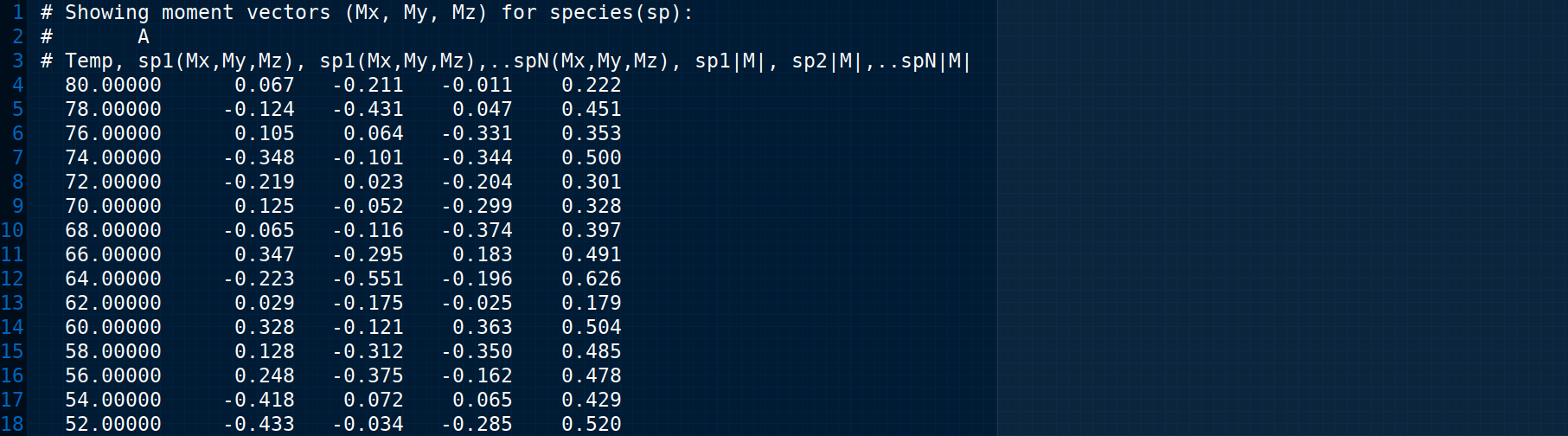}}
\caption{Illustrating the format of the `moment.dat' file evaluated for the ferromagnetic spin system arranged in a Simple Cubic lattice network. \newline\textit{After running the simulation, results can be obtained at location:}\\ /examples/SimpleCubic/333/data/moment.dat }
\label{mm}
\end{figure}

The $moment\_vector.dat$ file contains information about the temperature dependent net magnetic moment vector of the simulation box for each magnetic species. This allows users to visualize how the magnetic moment evolves on specific magnetic lattice networks as a function of temperature. The format of the $moment\_vector.dat$ file can be understood as follows:

\begin{enumerate}

\item \textit{The first line} contains the definition.  
\item \textit{The second line} provides the ordering of the relevant magnetic information.  
\item From the \textit{third line} onward, as shown in Figure \ref{mm}, the actual data are arranged in columns. The first column contains the temperature values. From the second column onward, the magnetic moment vectors (in the order $M_x$, $M_y$, $M_z$) are listed for each species, starting from the first species ($sp_1$) up to the $N^{th}$ species ($sp_N$). The subsequent columns, following the same order, present the magnitude of the magnetic moment for each species ($sp_i|M|$). It should be noted that the ordering of species here strictly follows the same order as listed in the $structure.vasp$ file.

\end{enumerate}


\newpage

\subsection{$magnetization.dat$}
\begin{figure}[h!]
\centering
\fbox{\includegraphics[width=0.9\textwidth]{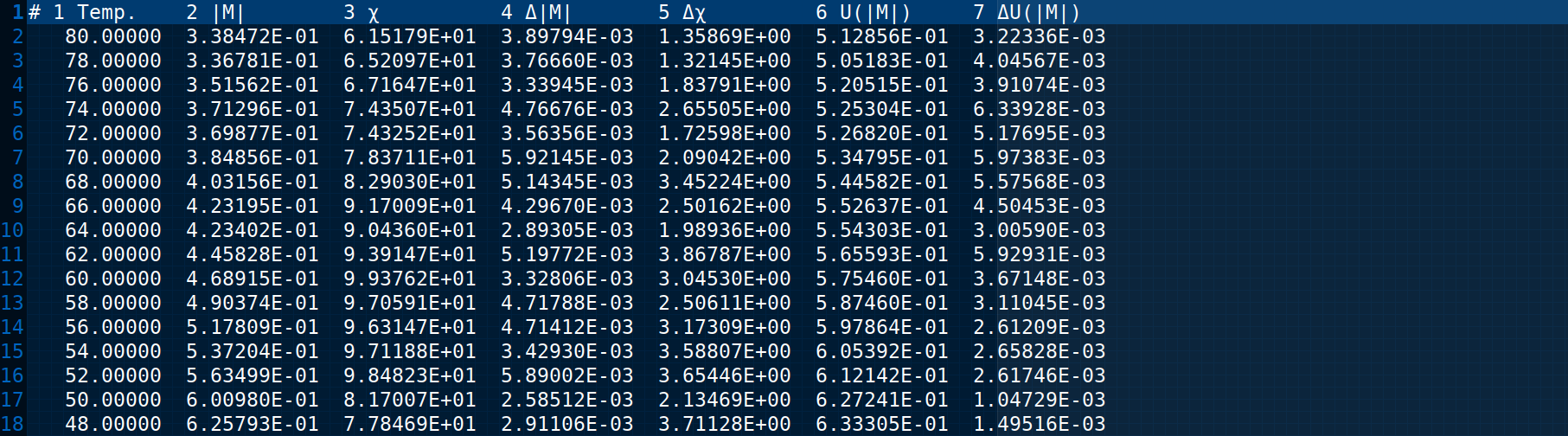}}
\caption{Shows the format of the `magnetization.dat' file obtained for a ferromagnetic spin system arranged on a simple cubic lattice network. \newline\textit{After running the simulation, results can be obtained at given location:}\\  /examples/SimpleCubic/333/data/magnetization.dat}
\label{magfile}
\end{figure}

This file contains the calculated magnetization as a function of temperature. The data are organized into seven columns (see Figure \ref{magfile}), which are as follows:

\begin{enumerate}
\item Temperature
\item Total average magnetization $|M|$, [eqn. \ref{mag}]
\item Susceptibility ($\chi$), [eqn. \ref{chi}]
\item Error in $|M|$, [eqn. \ref{se}]
\item Error in $\chi$, [eqn. \ref{se}]
\item Cumulant of magnetization, ($U_{Mag}$), [eqn. \ref{Cuml}]
\item Error in $U_{Mag}$, [eqn. \ref{se}]
\end{enumerate}


\newpage

\subsection{$energy.dat$}
This file contains the calculated energy dependent observables as a function of temperature. The data are organized into eight columns (see Figure \ref{magfile}), whose descriptions are as follows:

\begin{enumerate}
\item Temperature
\item Total internal energy ($E$), [eqn. \ref{eng}]
\item Specific heat ($C_V/k_B$), [eqn. \ref{sh}]
\item Error in $E$, [eqn. \ref{se}]
\item Error in $C_V/k_B$, [eqn. \ref{se}]
\item Cumulant of E, ($U_{E}$), [eqn. \ref{Cuml}]
\item Error in $U_{E}$, [eqn. \ref{se}]
\item Acceptance ratios (in \%) defined as how much MCS were accepted out of total MCS for generating new states.

\end{enumerate}

\begin{figure}[h!]
\centering
\fbox{\includegraphics[width=1.0\textwidth]{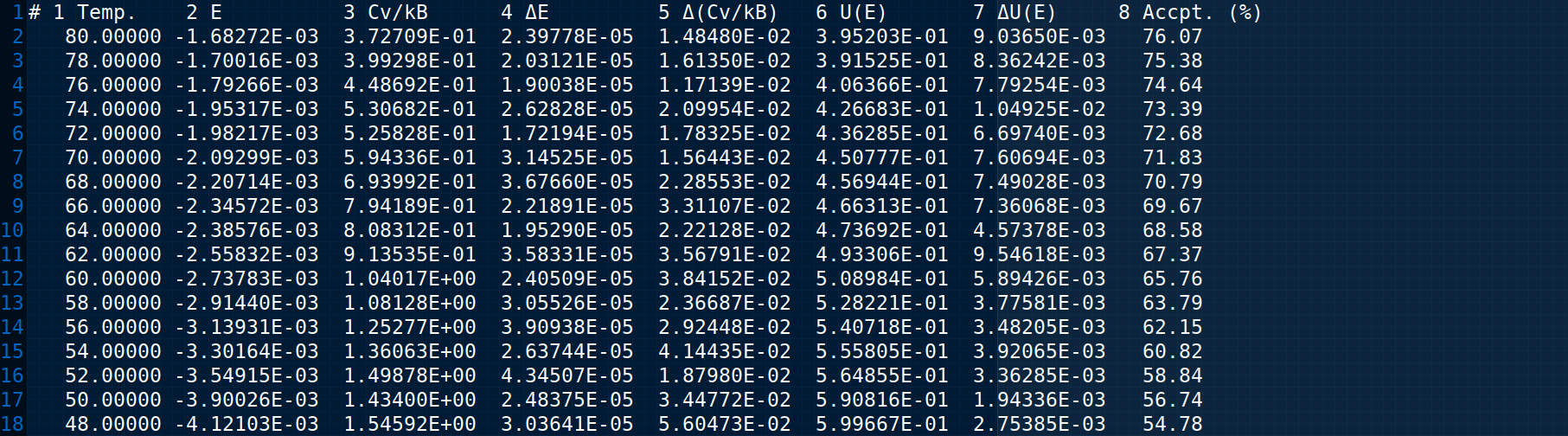}}
\caption{Preview of the `energy.dat' file obtained for a ferromagnetic spin system arranged on a simple cubic lattice network. \newline\textit{After running the simulation, results can be obtained at given location:}\\ /examples/SimpleCubic/333/data/energy.dat}
\label{engfile}
\end{figure}

\subsection{$graph.sh$}
File `$graph.sh$' contains the minimal `gnuplot' script to plot the obtained results for quick analysis.
\\

\fbox{\textit{COMMAND} $\Longrightarrow$ gnuplot graph.sh}
\\

\noindent Above command will generate a 2$\times$2 plot (file name: \textit{results\_Ether.png}) of total internal energy, specific heat, net magnetization, and susceptibility plots. Accordingly, user can also modify the $graph.sh$ to plot any other observable.


\newpage

\section{Utility tools}
\subsection{Vector}
The utility program `vector.f90' (/utility/Vector/vector.f90) can be used to visualize the final spin vectors obtained from $ETHER$ at a given temperature ($T$). Its main purpose is to plot the spin vectors on the surface of a unit sphere, with each vector originating from the center of the sphere. This visualization technique is particularly useful for large supercells, where examining the orientation of spin vectors directly can be difficult. Users must provide the spin file (`$<tt.tttt>$\textbf{spK.xsf}'), generated at temperature ($tt.tttt$), which is usually available in the \textbf{spin/} directory after completion of the simulation. The vector program reads this file and extracts the data required for visualization. The following are the commands to compile and run this tool:
\\

\fbox{\textit{COMMAND} $\Longrightarrow$ gfortran -o vector vector.f90} \vspace{0.5 cm}

\fbox{\textit{RUN} $\Longrightarrow$ ./vector}
\\

\noindent The descriptions of input/output files for this utility tool are as follows:

\subsubsection{$in\_vec.ether$}
This input file should be present before running the $vector$ program. The user must remove (if any) the older $<tt.tttt>spK.xsf$ file before running the $vector$ program. It seeks only two pieces of information whose descriptions are as follows,

\begin{enumerate}
\item Total number of ions which we want to include for vector plotting.
\item Label of ions, must be present in the `$structure.vasp$' file, separated by blank space.
\end{enumerate}
\noindent \textit{A typical example of this file is given below,}

\noindent ============================================

\noindent 3 \hspace{50 pt}! total number of ions to include, \textbf{First line} \\
A B C\hspace{1 cm}! their ionic symbols, \textbf{Second line}

\noindent ============================================

\subsubsection{$<$species\_name$>$\_vec}
Files with the name \textit{$<$species\_name$>$\_vec} will be generated after the execution of \textit{./vector}, such that label $<$species\_name$>$ will be replaced as per the species symbols mentioned in second line of \textit{in\_vec.ether} file e.g., \textit{A\_vec, B\_vec,} and \textit{C\_vec} if \textit{A, B, C} ions are listed in `\textit{structure.vasp}' file.

\subsubsection{$plot\_vector.sh$}

\begin{figure}[h!]
\centering
\includegraphics[width=0.6\textwidth]{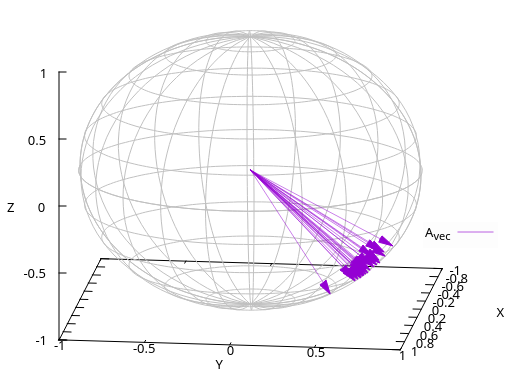}
\caption{Using the vector program, the ground spin state configurations observed at 5 Kelvin are plotted on the surface of unit sphere. \newline\textit{After running the simulation, results can be obtained at given location:}\\ /examples/SimpleCubic/333/spin/2.0000spK.xsf}
\label{vector}
\end{figure}

This is the \textit{gnuplot} script file and will be generated after the \textit{vector} program completion. The content of this file will also appear on the screen, and the user may directly copy this content and paste it directly into the \textit{gnuplot} terminal for 3D visualization (Figure \ref{vector}) of vectors on the unit sphere. 

\subsubsection{$sp\_vector.xsf$}
This file is the exact copy of the provided $<tt.tttt>spK.xsf$ file and is used by $vector$ program internally.

\subsection{Structure Factor}
Magnetic structures (commensurate or incommensurate) and their associated magnetic wave vectors ($q$), are crucial for understanding non collinear magnetic behavior. Evaluating the structure factor ($S_f$), defined as the thermal average of the Fourier transformed spin-spin correlation function, can help identify complex magnetic phase transitions as a function of temperature.

Using the utility tool `$structure\_factor.f90$' (/utility/Vector/structure\_factor.f90), users can generate a heat map of $S_f$ as a function of $q$ at different temperatures using the following relation:

\begin{equation}\label{sf1}
S_f(q) = \frac{1}{N}\sum_{r,\,r'} \left <\,S_{r} S_{r'} \right > e^{iq(r-r')} 
\end{equation}

\begin{figure}[ht!]
    \centering
    \includegraphics[width=0.5\textwidth]{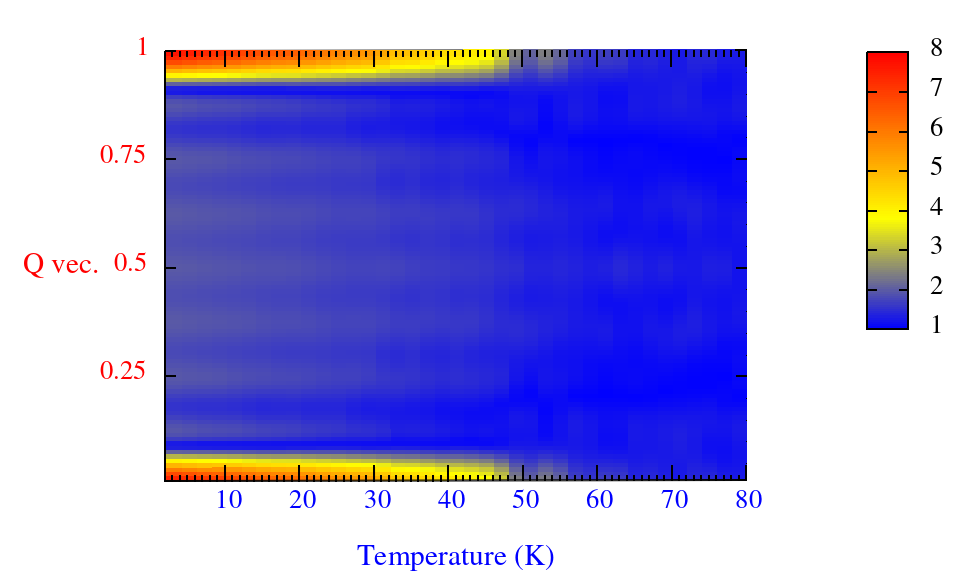}\includegraphics[width=0.5\textwidth]{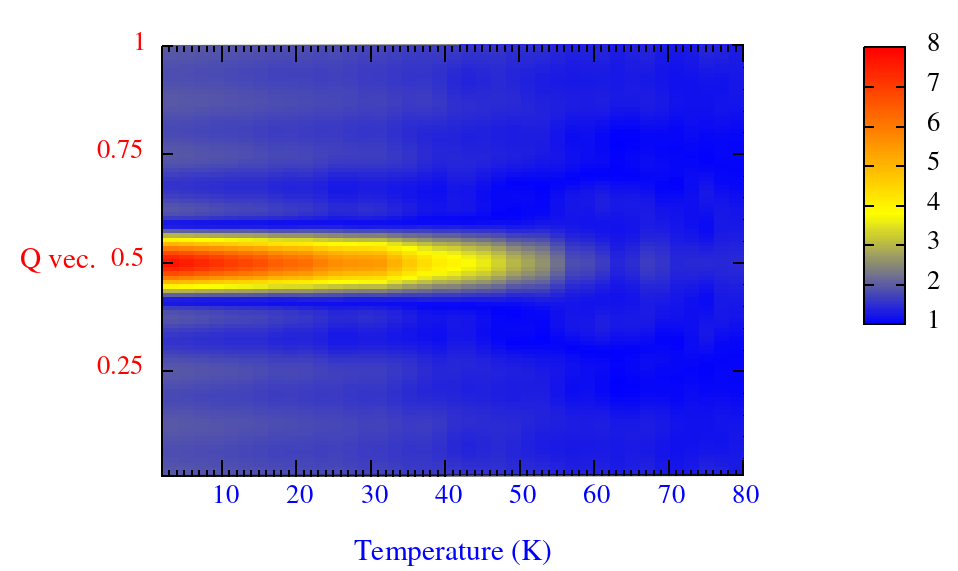}
\caption{The structure factor heat map as a function of temperature obtained using $structure\_factor.f90$ are demonstrated here. The plots shown in the left and right panels correspond to the ferromagnetic and antiferromagnetic cases, respectively. \newline\textit{Evaluated for the $gss.dat$ at location:}\newline/examples/SimpleCubic/FM/data/ and /examples/SimpleCubic/AFM/data}
    \label{sf2}
\end{figure}

Since $q$ is proportional to $\frac{1}{L}$, where $L$ is the lattice parameter, spins arranged in a simple cubic lattice network with ferromagnetic and antiferromagnetic interactions exhibit maxima in $S_f$ at integer multiples of $q$ and $\frac{q}{2}$, respectively. Figure \ref{sf2} clearly validates these theoretical expectations.
\\

\fbox{\textit{COMMAND} $\Longrightarrow$ gfortran -o sf structure\_factor.f90} \vspace{0.5 cm}

\fbox{\textit{RUN} $\Longrightarrow$ ./sf}
\\

$structure\_factor.f90$ requires two input files: `$in\_SF.ether$' and ground spin states ($gss.dat$) for structure factor evaluations.

\newpage

\subsubsection{$in\_SF.ether$}
 The file $in\_SF.ether$ contains the necessary information which are as follows,
 
\begin{enumerate}
\item Direction along which we want to calculate the structure factor.
\item Latency (in degrees) used to evaluate the spins along the specified $directions$.
\item Range of $q$ values: $q_{final}$, $q_{initial}$, $dq$.
\item Scaling factor used to scale the lattice parameters.
\end{enumerate}

\noindent \textit{A typical example of this file is given below,}

\noindent ============================================\\
\noindent 0 0 1 \hspace{50 pt}! Direction\\
\noindent 1\hspace{71 pt}! Least latency in pattern\\
\noindent 2 0.01 0.01\hspace{23 pt}! Magnetic wave vector: q(to), q(from), q(step size)\\
\noindent 0.5 0.5 0.5\hspace{26 pt}! (a, b, c), use to scale the Lattice parameters (a*L, b*L, c*L)

\noindent ============================================

\subsubsection{$plot\_sf.sh$}
This is an output file generated by $structure\_factor.f90$ that contains the required $gnuplot$ script to generate a heat map plot and save it as $figSF.png$. Note that this script uses the output file $outputSF.dat$ produced by the $structure\_factor.f90$ tool. To generate the plot, execute the following command in the terminal:
\\

\fbox{\textit{COMMAND} $\Longrightarrow$ gnuplot plot\_sf.sh}
\\

\newpage
\section{ETHER Benchmarking}
We considered three prototypical lattice systems namely simple cubic, pyrochlore, and stacked triangular lattices, each representing distinct types of magnetic interactions and ordering phenomena. 
These systems provide useful test cases for validating Monte Carlo implementations because their critical behaviors and thermodynamic properties have been extensively studied in the literature.
We finally benchmarked our MC results against results reported in the literature \cite{peczak1991,angle2001, passos2016,Bunker1993}.
  
\subsection {Simple Cubic Lattice}
We performed Monte Carlo simulations over the temperature range 1.0 to 1.8 with a temperature step of 0.01 to investigate the critical coupling constant  (K$_C$ = $\frac{k_BT_C}{J}$) for the simple cubic lattice with a ferromagnetic nearest neighbour (NN) interaction (J $<$ 0), within the classical Heisenberg model given by:

\begin{equation}\label{H_scl}
H = \sum_{i>j}J \vec S_i\cdot \vec S_j
\end{equation}

\begin{figure}[ht!]
    \centering
    \includegraphics[width=0.25\textwidth]{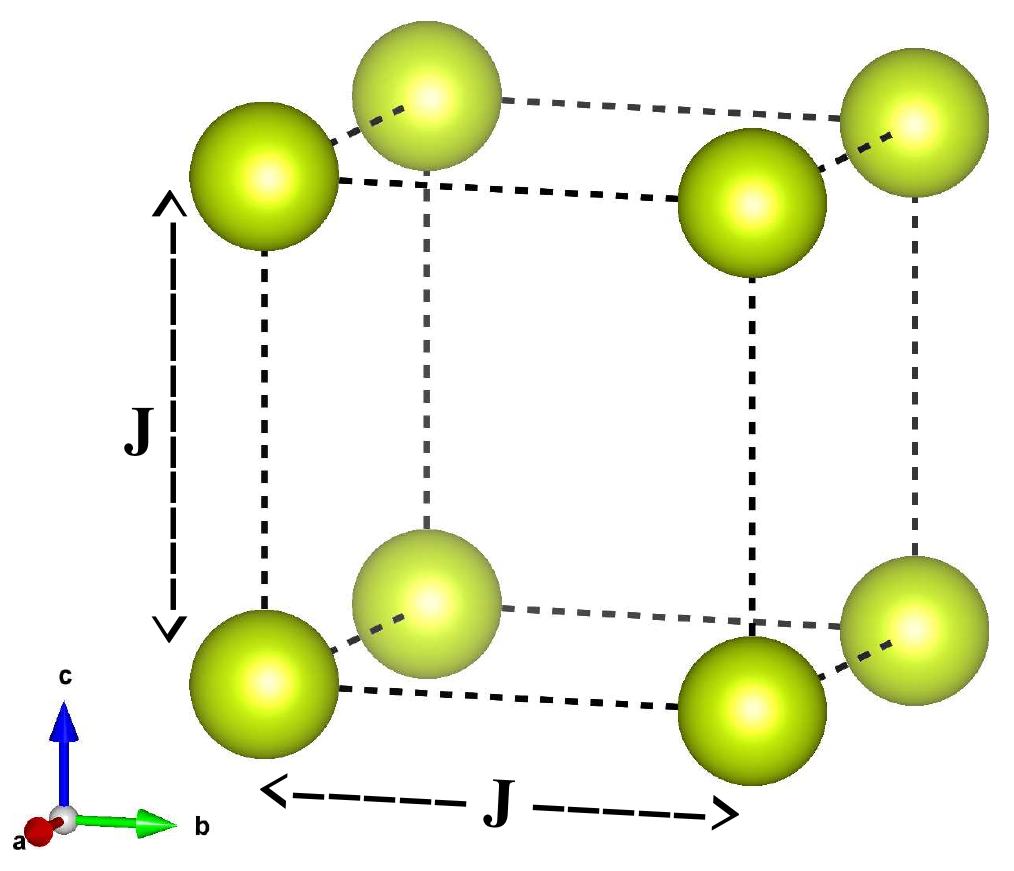}\includegraphics[width=0.75\textwidth]{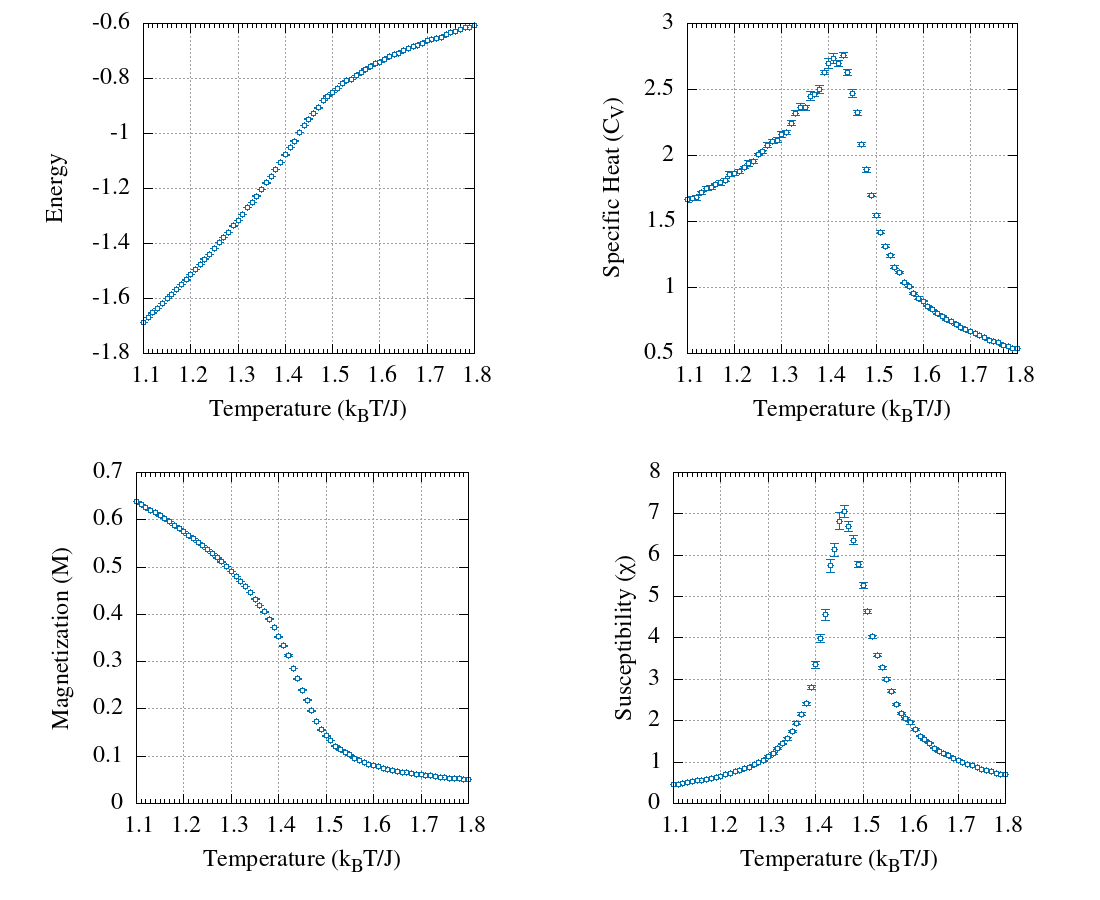}
    \caption{Plot showing the MC results for simple cubic lattice. The calculated observables per unit site such as energy, specific heat, magnetization and susceptibility is plotted w.r.t. the temperature $\frac{k_BT_C}{J}$. The calculations can be obtained in the file location : /examples/SimpleCubic/141414/ }
    \label{Sh_scl}
\end{figure}

Figure \ref{Sh_scl} presents our estimated results for a lattice size of $14\times14\times14$. Using the Metropolis algorithm in combination with the over-relaxation method, the specific heat and susceptibility curves exhibit their highest peaks at approximately $K_C \approx 1.44$, which agrees well with the reported value $K_C = 1.4432$ \cite{peczak1991}. Even closer agreement with the reported value (up to four significant digits) can be achieved by performing MC simulations with a smaller temperature step, increasing the number of MCS steps, and using larger lattice sizes.

\subsection {Pyrochlore Lattice}

\begin{figure}[ht!]
    \centering
    \includegraphics[width=0.25\textwidth]{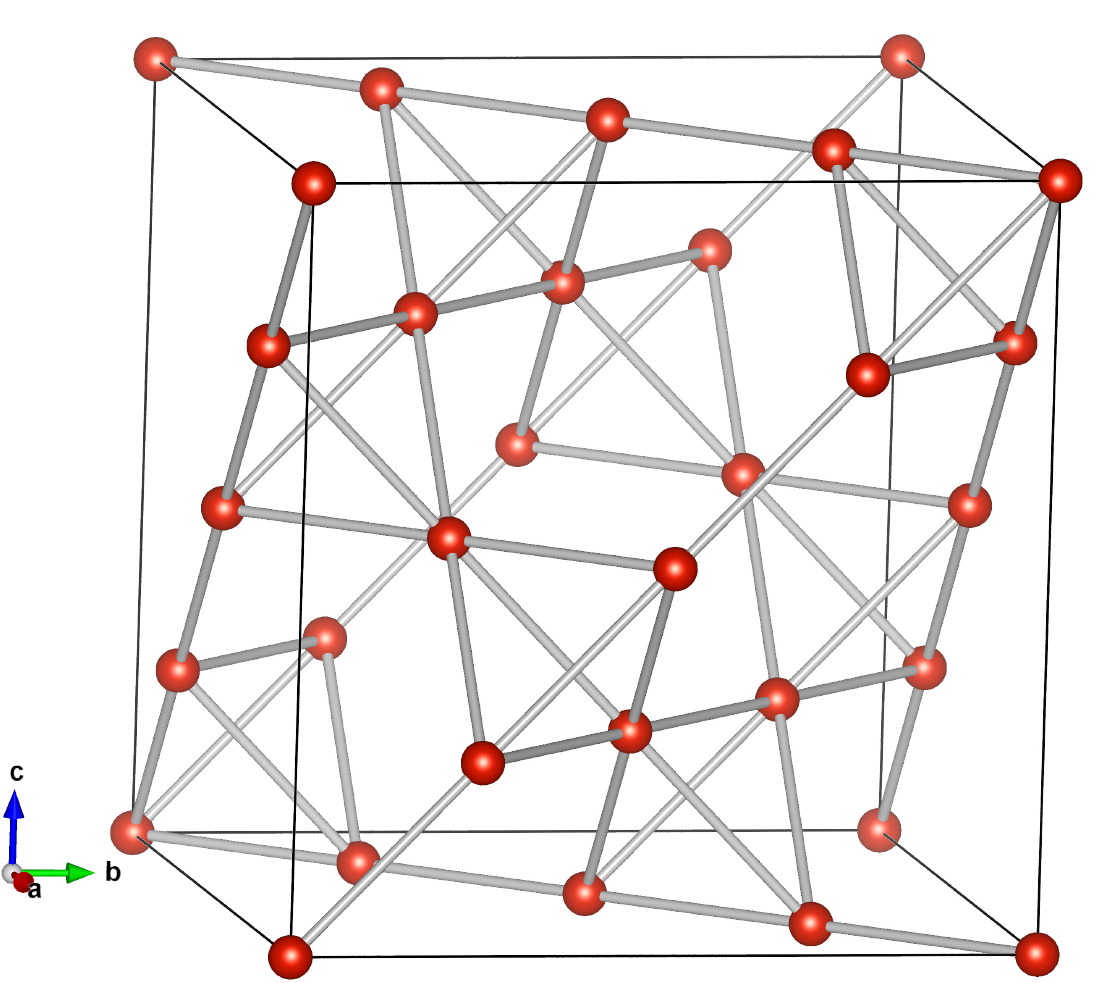}\includegraphics[width=0.75\textwidth]{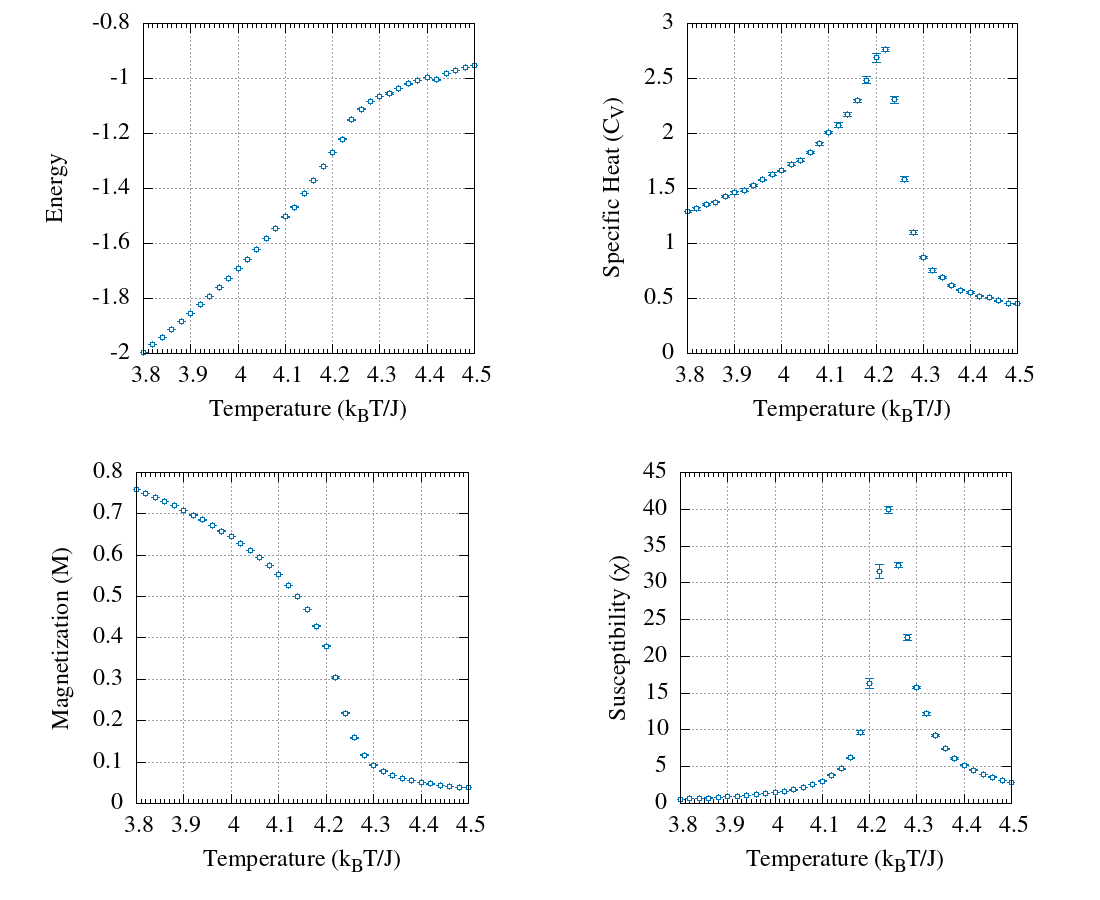}
    \caption{MC results of Pyrochlore lattice. The calculations can be obtained in the file location : /examples/Pyrochlore}
    \label{pcl}
\end{figure}

In the second test, we modeled the pyrochlore lattice network (Figure \ref{pcl}) with ferromagnetic NN interactions using the classical Ising model (equation \ref{H_pcl}) and investigated the critical temperature ($K_C$). The Hamiltonian can be written as : 

\begin{equation}\label{H_pcl}
H = \sum_{i>j}J  S_i^z  S_j^z
\end{equation}

Since the pyrochlore network with ferromagnetic interaction exhibits only a single magnetic phase transition, we tested our ETHER code with the reported critical temperature value ($K_C = 4.0$) \cite{Bramwell1998} for comparison. We performed extensive MC calculations within the temperature ($k_BT/J$) range of 3.8 to 4.5 with a temperature step of 0.02. Figure \ref{pcl} shows the calculated observables as a function of temperature for a lattice size of $10\times10\times10$ supercell. A peak is observed around $k_BT/J \approx 4.2$, which is close to 4 and also demonstrates the consistency of our MC results with previously reported values (4.347826 \cite{angle2001}, 4.213893 \cite{passos2016}).

\subsection {Triangular Lattice}
\begin{figure}[!h]
    \centering
    \includegraphics[width=0.25\textwidth]{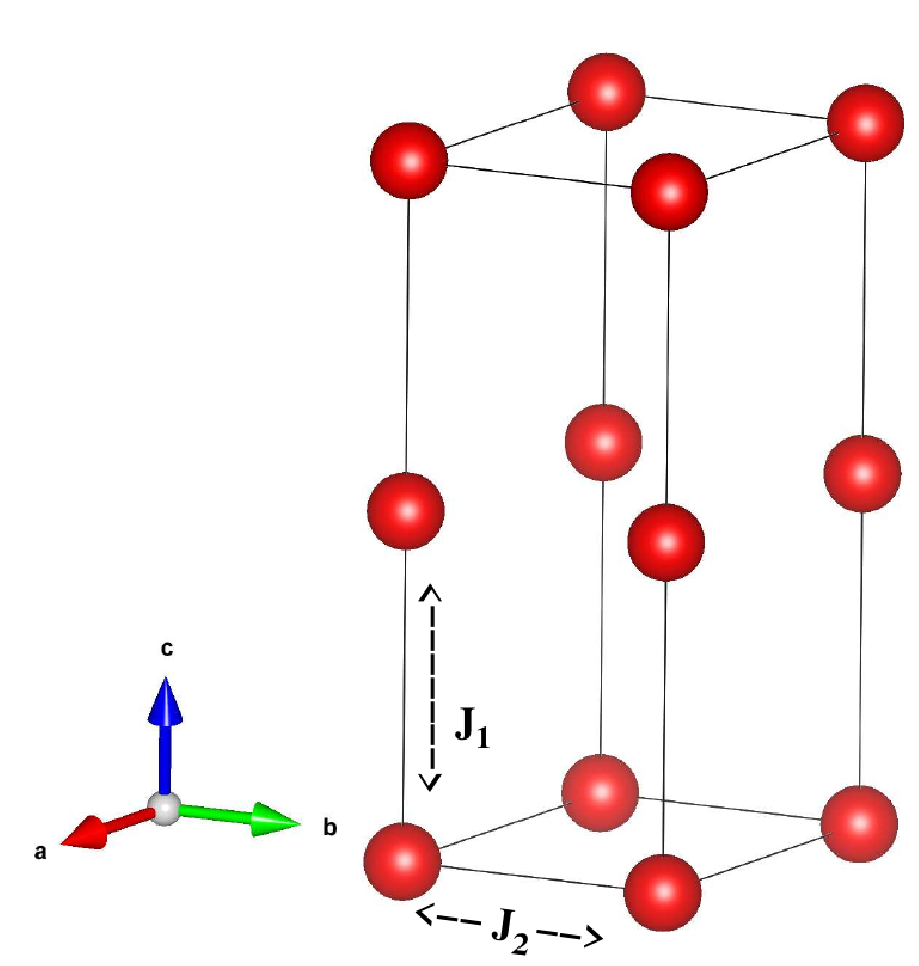}\includegraphics[width=0.75\textwidth]{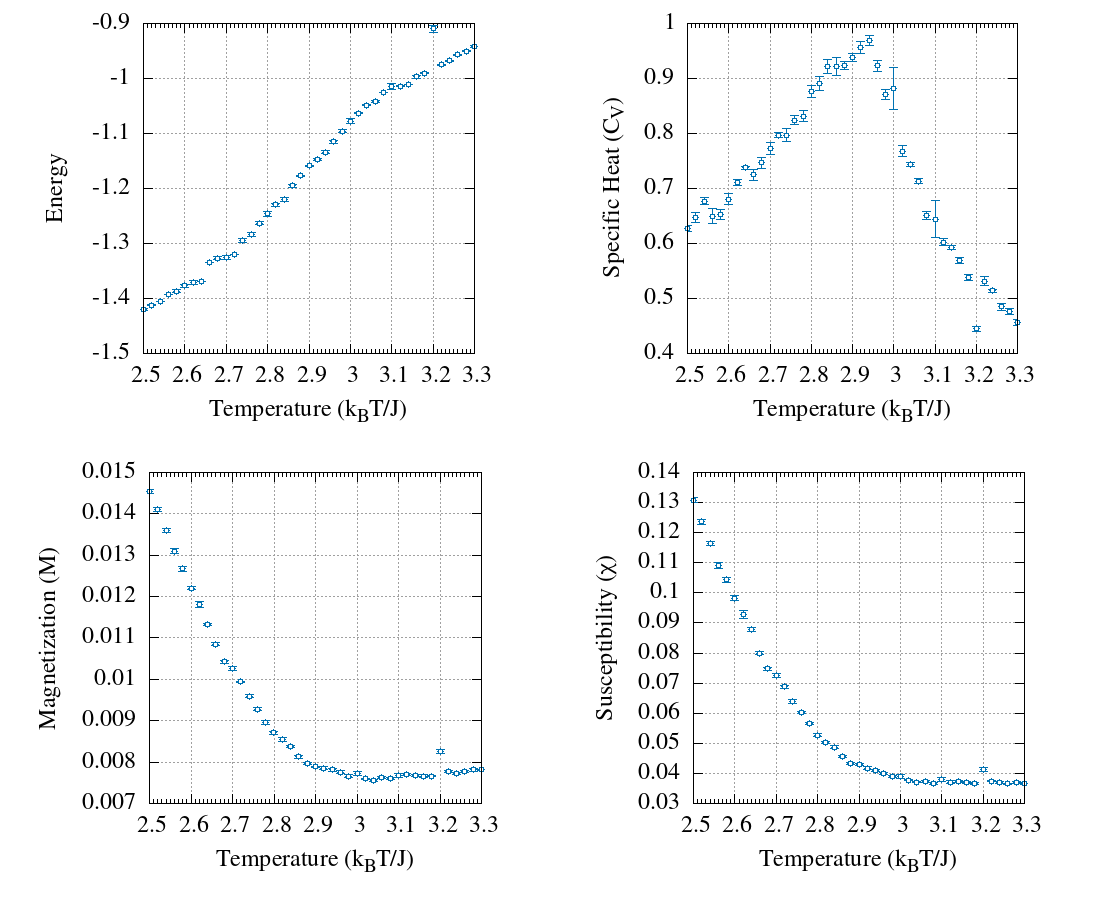}
    \caption{MC results of stacked triangular lattice. The calculations can be obtained in the file location : /examples/Triangular}
    \label{stl}
\end{figure}

Finally, in this section, we modelled a stacked triangular lattice (Figure \ref{stl}) with out-of-plane (along $c$) ferromagnetic interactions ($J_1 < 0$) and in-plane antiferromagnetic ($J_2 > 0$) using the Hamiltonian given as,

\begin{equation}\label{H_stl}
H = \sum_{i>k}J_1\,  S_i^z  S_k^z + \sum_{i>j}J_2\,  S_i^z  S_j^z 
\end{equation}

Equation \ref{H_stl} is adopted from the Ising model, where the site index $k$ runs along the $c$ direction. $J_1$ and $J_2$ represent the interactions between the nearest neighbour ions, as shown in Figure 12. In our MC calculations, we chose $|J_1| = |J_2| = |J|$. A peak at 2.94 ($k_B T/J$) in the specific heat profile (Figure \ref{stl}), calculated for a lattice size of $12\times12\times12$, is very close to the previous MC result reported by Bunker et al. \cite{Bunker1993}, where they obtained a critical temperature of 2.92.

\section{Acknowledgement}
I acknowledge Prof. Tulika Maitra, Indian Institute of Technology Roorkee (IITR) and Dr. Jyoti Krishna, IITR for the valuable discussions during my PhD. 

\clearpage

\bibliography{ref_ETHER}
\bibliographystyle{mybib}

\end{document}